# Nodal and nematic superconducting phases in NbSe$_2$ monolayers from competing superconducting channels

Chang-woo Cho[1], Jian Lyu[1,2], Liheng An[1], Tianyi Han[1], Kwan To Lo[1], Cheuk Yin Ng[1], Jiaqi Hu[1], Yuxiang Gao[1], Gaomin Li[2], Mingyuan Huang[2], Ning Wang[1], Jörg Schmalian[3], and Rolf Lortz[1,*]

[1]Department of Physics, The Hong Kong University of Science and Technology, Clear Water Bay, Kowloon, Hong Kong.

[2]Department of Physics, Southern University of Science and Technology, 1088 Xueyuan Road, Nanshan District, Shenzhen, Guangdong Province, China.

[3]Institute for Theory of Condensed Matter and Institute for Quantum Materials and Technologies, Karlsruhe Institute of Technology, Karlsruhe, Germany.

Transition metal dichalcogenides like 2H-NbSe$_2$ in their two-dimensional (2D) form exhibit Ising superconductivity with the quasiparticle spins are firmly pinned in the direction perpendicular to the basal plane. This enables them to withstand exceptionally high magnetic fields beyond the Pauli limit for superconductivity. Using field-angle-resolved magnetoresistance experiments for fields rotated in the basal plane we investigate the field-angle dependence of the upper critical field ($H_{c2}$), which directly reflects the symmetry of the superconducting order parameter. We observe a six-fold nodal symmetry superposed on a two-fold symmetry. This agrees with theoretical predictions of a nodal topological superconducting phase near $H_{c2}$, together with a nematic superconducting state. We demonstrate that in NbSe$_2$ such unconventional superconducting states can arise from the presence of several competing superconducting channels.

2H-NbSe$_2$ is a layered transition metal dichalcogenide (TMDC) superconductor with bulk critical temperature $T_c$ = 7.2 K, which remains superconducting in monolayer form [1,2]. The layers have honeycomb lattice structure with broken A-B sub-lattice and inversion symmetry and heavy transition metal atoms causing strong spin-orbit coupling (SOC) [3-5]. The electrons in the layer experience in-plane electric fields from a broken in-plane mirror symmetry. This leads to an Ising SOC field [6], which pins the electron spins to the out-of-plane direction [7-9]. Cooper pairs are thus not affected by the Pauli paramagnetic limit and can exist in very high magnetic fields parallel to the layers [1,6]. Ising superconductivity in monolayer NbSe$_2$ has been predicted to feature a topological superconducting phase with 6 pairs of point nodes on the lines connecting the Γ and the M points in the Brillouin zone [10-12], protected by time-reversal symmetry and connected by Majorana arcs [13-15]. In this Letter, we investigate the pairing gap symmetry of NbSe$_2$ monolayers by field-angle-resolved electric transport measurements in strictly parallel fields. The angular $H_{c2}$ dependence directly reflects the superconducting order parameter symmetry [16,17]. We find that the in-plane field drives the superconductor into an unconventional state, with the predicted six-fold nodal symmetry [10], and a nematic phase [18] consisting of a primary single-component order parameter competing with a sub-leading unconventional two-component order parameter.

*Corresponding author: lortz@ust.hk

In total three monolayer samples have been studied with consistent results in comparison to thicker samples in which the effects were much weaker. Here we present results from a representative monolayer (Sample #1) while data from other samples can be found in Ref. [19]. The 2D NbSe$_2$ samples were fabricated using the standard exfoliation method. The device was prepared by depositing a hexagonal boron-nitride layer on a SiO$_2$/Si substrate with 15 nm deep grooves cut by plasma etching and filled with Au to form non-protruding bottom electrodes. NbSe$_2$ flakes were exfoliated onto PMMA and transferred onto the substrate, keeping the PMMA as protective layer. Fig. 1a shows an optical image of the device. Micro-Raman spectroscopy was used to confirm the presence of a monolayer region between the terminals (inset) [19] . Electrical magneto transport measurements were performed with a standard four-probe AC method. The devices were mounted on a piezo rotatory stage with its axis perpendicular to the field. Exact parallel field alignment of the NbSe$_2$ layers was secured by a goniometer. Fig. 1a shows an optical image of the device where the monolayer flake is connected by 5 terminals. Fig. 1b shows the zero-field superconducting transition with $T_c$ characteristic for a monolayer [1]. In Fig. 2a, we plot 12-T resistance data for Sample #1 for different in-plane field angles $\Phi$ (Fig. 1b). A significant angular variation is observed. We marked characteristic temperatures at which the resistance reaches a certain value at the midpoint (50% of normal state resistance $R_N$) and at 90% $R_N$ of the superconducting transition by stars. Fig. 2b & c show the $\Phi$ dependence of these characteristic temperatures. At 90% $R_N$, the angular dependence shows a pronounced 6-fold variation with sharp kinks each 60º superimposed by a 2-fold variation. At 50% $R_N$, the kinks are less evident, and the 2-fold symmetry dominates. A very similar behavior with two distinct field regions, in which there is 6-fold nodal and 2-fold symmetry, respectively, was observed in our other monolayer devices Sample #2 & 3 [19].

To illustrate the $\phi$ variations of the characteristic temperatures, we used Eq. 1 as a fitting function. The first term describes a 2-fold nematic symmetry [20] to which we add a $|\cos(3\phi + \phi_6)|$ term to model a 6-fold nodal symmetry with its sharp peaks caused by the absolute value of the cosine function.

$$H_{c2}(\phi) = \frac{A_2}{\sqrt{\cos^2(\phi+\phi_2)+\Gamma^2 \sin^2(\phi+\phi_2)}} + A_6|\cos(3\phi + \phi_6)| \qquad (1)$$

$\Gamma$ is the anisotropy parameter, $A_i$ and $\phi_i$ ($i$ = 2, 6) indicate the gap amplitude and phase. Note that while these equations describe the anisotropy for $H_{c2}(T,\phi)$, our temperature dependent resistance data provides $T_c(H,\phi)$, which is directly related to $H_{c2}(T,\phi)$ and follows the same $\phi$-dependence. The fitting functions are included as lines in Fig. 2b & c. A nodal 6-fold gap function can describe the kinks near the normal state boundary very well. In addition, a 2-fold variation must be considered.

We attribute the 6-fold sharp kink-like dependence of $H_{c2}$ to the theoretically predicted nodal superconducting phase [10]. The additional 2-fold variation contradicts the trifold crystalline in-plane symmetry and can be regarded as a nematic superconductivity similar to doped $Bi_2Se_3$ [18]. Our studies on thicker samples revealed that the strong anisotropies are exclusively found in $NbSe_2$ monolayers, while multilayers we measured had much weaker anisotropies [19].

**Discussion**

The $NbSe_2$ monolayers are subject to a strong Ising SOC and a Zeeman field in the plane. The broken inversion symmetry in the crystal symmetry causes a mixture of spin singlet and spin triplet pairing potential [18,20,21]. The interplay between the Zeeman field, SOC and eventually disorder is regarded to be crucial for tuning the coupling among the different pairing channels [22]. We observe the kinks of the 6-fold rotational symmetry for all monolayer samples near the normal state boundary. At lower temperatures and fields, the kinks become less evident (Sample #1) or even absent (Sample #2 & 3) [19], and the field angle dependence is dominated by the 2-fold symmetry. This suggest that there is a region near $H_{c2}(T)$ with a 6-fold nodal superconducting gap symmetry. This is consistent with the theoretical predictions [10], where a 6-fold nodal topological superconducting phase is induced by high parallel magnetic fields in monolayer $NbSe_2$. The nodes represent 6 pairs of point nodes on the two Γ pockets along the three Γ-M lines in the Brillouin zone, which are protected by the time-reversal-like symmetry [21]. Ising SOC is absent along these directions, so the Zeeman energy can exceed the pairing energy of the spin-singlet Cooper pairs, thus closing the gap and forming point nodes [10]. It has been shown [12] that there are two instabilities in singlet and triplet interaction channels leading to a crystalline topological superconducting phase, which involves the nodal phase [10]. Such a nodal crystalline topological superconducting phase has particle-hole symmetry and an anti-unitary time-reversal like symmetry, the latter being a composition of time-reversal symmetry and a reflection with respect to the *xy*- plane [12]. The point nodes in the predicted topological superconducting phase should be connected by Majorana flat bands [10], this stimulates more direct spectroscopic experiments.

In Fig. 3 we summarize our results in an *H-T* phase diagram. The stars indicate the regions in which we observe evidence for the nodal phase. Our transport data do not provide information in the zero-resistance region, but the additional two-fold rotational symmetry is observed wherever the resistance remains finite, and, most importantly, is reflected in a field-angle dependence of the normal state boundary (see inset of Fig. 2a). This can be plausibly explained by a nematic superconducting state.

Nematic superconductivity means that a rotational symmetry breaking occurs, which is closely related to the onset of superconductivity. Artefacts due to misalignment of the sample surface with respect to the axis of rotation can be excluded [19]. We have checked the parallel alignment carefully at the orientations where the extreme values occur. Furthermore, in none of our devices a correlation between the current direction and the observed two-fold symmetry seems to exist [19], which excludes anisotropies due to dissipation from vortices [23]. In the inset of Fig. 2a, it can be seen that the 12T-resistance data coincides for 0° and 180°, but the 90°-data is shifted to lower temperature without changing its shape, which demonstrates that the anisotropy originates from a true $H_{c2}$ anisotropy and not only from an in-plane variation of the sheet resistance. For Sample #1 the orientation of the 2-fold symmetry varies even slightly as a function of temperature, and we found that in the region between contact 1 and 2 the orientation is different, as evidenced by a series of 2-probe measurements, which suggests the presence of different nematic domains [19].

In the following we present a scenario, which naturally accounts for the observations and discuss the details of the phase diagram (Fig. 3a). It implies that there are several competing superconducting channels in $NbSe_2$: a dominant single-component (likely *s*-wave) pairing state and a sub-leading, unconventional two-component order parameter, which leads to nematicity because of the symmetry-permitted nonlinear coupling between both pairing states (Fig. 3b).

In a nematic superconductor, the onset of superconductivity not only breaks the global $U(1)$ symmetry of the pairing state, but also a discrete rotational symmetry of the crystal lattice, which can even occur above the actual superconducting phase transition [24-26]. For such a nematic state to exist, the superconducting order parameter must transform according to a higher-dimensional irreducible representation of the point group. The point group of free-standing monolayer NbSe$_2$ is $D_{3h}$. On a substrate, the horizontal mirror plane disappears, and the resulting group becomes $C_{3v}$. In $C_{3v}$ there is a 2D irreducible representation $E$ with leading polynomials $(k_x, k_y)$, or $(k_x^2 - k_y^2, k_x k_y)$. A superconductor that orders according to this $E$ representation would then be characterized by

$$\Delta_{\alpha\beta}(k) = \psi_1\big(\Delta_{x^2-y^2}(\boldsymbol{k}) + \boldsymbol{d}_x(\boldsymbol{k}) \cdot \boldsymbol{\sigma}\big)i\sigma_y + \psi_2\big(\Delta_{xy}(\boldsymbol{k}) + \boldsymbol{d}_y(\boldsymbol{k}) \cdot \boldsymbol{\sigma}\big)i\sigma_y \qquad (3)$$

Here $\Delta_{\alpha\beta}(\boldsymbol{k}) = \langle c_{\boldsymbol{k}\alpha} c_{-\boldsymbol{k}\beta} \rangle$ describes the Cooper pair with crystal momentum $\boldsymbol{k}$ and spin $\alpha, \beta$. $\Delta_{x^2-y^2}(\boldsymbol{k}) \propto \cos k_x - \cos k_y$ and $\Delta_{xy}(\boldsymbol{k}) \propto \sin k_x \sin k_y$ describe singlet pairing amplitudes while the $\boldsymbol{d}_{x,y}(\boldsymbol{k})$ describe the triplet component, which is allowed given the broken inversion symmetry at the interface. The pairing is then characterized by the two-component order parameter $\psi = (\psi_1, \psi_2)$. A nematic state has the helical form $(\psi_1, \psi_2) \propto (\cos\theta, \sin\theta)$. An immediate concern of such a pairing state is that one would expect pairing states in bulk NbSe$_2$ that naturally merge with the $E$ representation on the surface. After all, the bulk and surface transition temperatures are rather comparable. These pairing states would then be either of $E_{2g}$ or $E_{1u}$ symmetry with the bulk point group $D_{6h}$. There seems to be no evidence for such behavior in bulk NbSe$_2$. Another, arguably more fundamental objection against primary nematic superconductivity in monolayer NbSe$_2$ follows from the analysis of Ref. 27, where monolayers with broken inversion symmetry were considered in the limit where the spin splitting due to the inversion symmetry breaking is larger than the superconducting gap, a condition which is satisfied. It has been shown that superconductivity with higher-dimensional irreducible representations, such as $E$, always breaks time reversal symmetry, i.e. $(\psi_1, \psi_2) \propto (1, \pm i)$, instead of being nematic.

A natural explanation for the observed behavior is that the primary superconducting order parameter $\varphi$ is a single-component degree of freedom. However, in addition to this primary superconducting order parameter, pairing in the *E*-symmetry channel, i.e. with the pairing wave function given in Eq. 3 with the two-component order parameter $\psi = (\psi_1, \psi_2)$, is a close contender. Then, the symmetry properties at the interface allow a phase transition $T_{nem} < T_c$ at zero field (Fig. 3a), where $\psi$ becomes finite and enters a nematic state (Fig. 3b). This nematic order is a direct consequence of the nonlinear coupling between the two almost degenerated order parameters. The most conservative choice would be *s*-pairing in the $A_1$ symmetry. Then a coupling of the type

$$f_{int} = \frac{g}{4}[\varphi^*(\psi_1|\psi_1|^2 - \psi_2\psi_1^*\psi_2 - 2\psi_1|\psi_2|^2) + h.c.] \quad (4)$$

is symmetry allowed and induces a nematic state which is either $\psi^{(1)} = \psi_0(1,0)$, $\psi^{(2)} = \psi_0\left(-\frac{1}{2}, \frac{\sqrt{3}}{2}\right)$, or $\psi^{(3)} = \psi_0\left(-\frac{1}{2}, -\frac{\sqrt{3}}{2}\right)$ with amplitude $\psi_0 \propto \varphi$ below a first order transition at $T_{nem}$. A similar behavior occurs if $\varphi$ is odd under vertical mirror reflections, i.e. an $A_2$ order parameter, which yields instead

$$f_{int} = \frac{g}{4}[\varphi^*(\psi_2|\psi_2|^2 - \psi_1\psi_2^*\psi_1 - 2\psi_1\psi_1^*\psi_2) + h.c.] \quad (5)$$

with similar nematic order. These nematic states occur for a sufficiently large coupling constant *g*, even though the secondary order parameter $\psi$ prefers, on its own, a chiral state with time-reversal symmetry breaking [27]. Interestingly, a coupling term like Eq. 4 has recently been discussed in the context of nematic superconductivity in twisted bilayer graphene [28].

If there are two almost degenerated order parameters that transform simultaneously under $A_1$ and *E*, an in-plane magnetic field plays an important role in mixing these two states below $T_c$ (Fig. 3b). This is a consequence of the magnetic field induced bi-linear coupling between the two order parameters, which induces the nematic state anywhere below $T_c$. Only directly at the transition temperature are nematic effects absent, which explains the observed 6-fold symmetry. This magnetic field coupling term is even under time reversal, is gauge invariant and compatible with all point symmetries. Within a microscopic theory, the physics underlying this coupling was recently discussed in Ref. 22.

To summarize, while the 6-fold symmetry observed agrees perfectly with the theoretical prediction of a field-induced topological superconducting phase [10], the 2-fold symmetry reveals that this material is characterized by a primary order parameter, likely of *s*-wave nature, and a two-component close competitor that transforms according to a non-trivial symmetry. The nonlinear coupling between the two order parameters induces a nematic state at $T_{\text{nem}} < T_{\text{c}}$ and pins a 2-fold symmetry axis in the system. An in-plane magnetic field induces nematic order at temperatures even above $T_{\text{nem}}$, all the way up to $T_{\text{c}}$. A role of strain in determining the pinning direction of the two-fold anisotropy is likely. In doped $Bi_2Se_3$ [26], intrinsic strain from a separate nematic transition above $T_{\text{c}}$ is crucial in aligning the direction of the anisotropic gap along a particular in-plane direction. For 2D $NbSe_2$, the strain field is likely extrinsic from the exfoliation process. It is unlikely that strain alone can explain the 2-fold anisotropy with a critical field variation up to 3 T. This requires an unconventional order parameter.

After completion of this work, we became aware of a preprint that reported a similar two-fold breaking of rotational symmetry but in few-layer $NbSe_2$ [29], which further confirms our experimental results. No evidence for a nodal phase was reported.

**Acknowledgements**

We thank U. Lampe for technical support and X. Dai, K.T. Law and R. Fernandez for enlightening discussions. This work was supported by grants from the Research Grants Council of the Hong Kong Special Administrative Region, China (GRF-16302018, GRF-16300717, C6025-19G-A, SBI17SC14) and by the German Research Foundation (DFG) through CRC TRR 288 "ElastoQMat", project B01.

**Figure Captions**

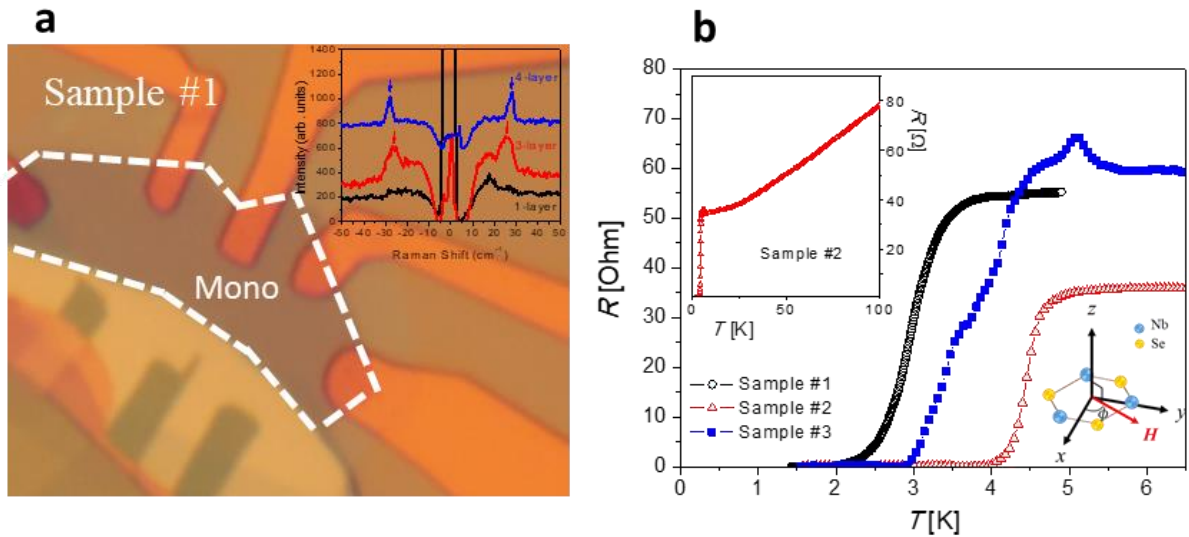

FIG. 1. a) Optical image of Sample #1 showing the monolayer region under investigation in the electrical transport experiments. The inset shows local Raman spectra at different positions used to identify the monolayer region. The spectra were shifted vertically. b) Low temperature zero-field electrical resistance of the 3 monolayer devices studied. The left inset shows data up to 100 K. The right inset defines the in-plane field orientation in the honeycomb lattice with respect to the $x$-axis (angle $\phi$).

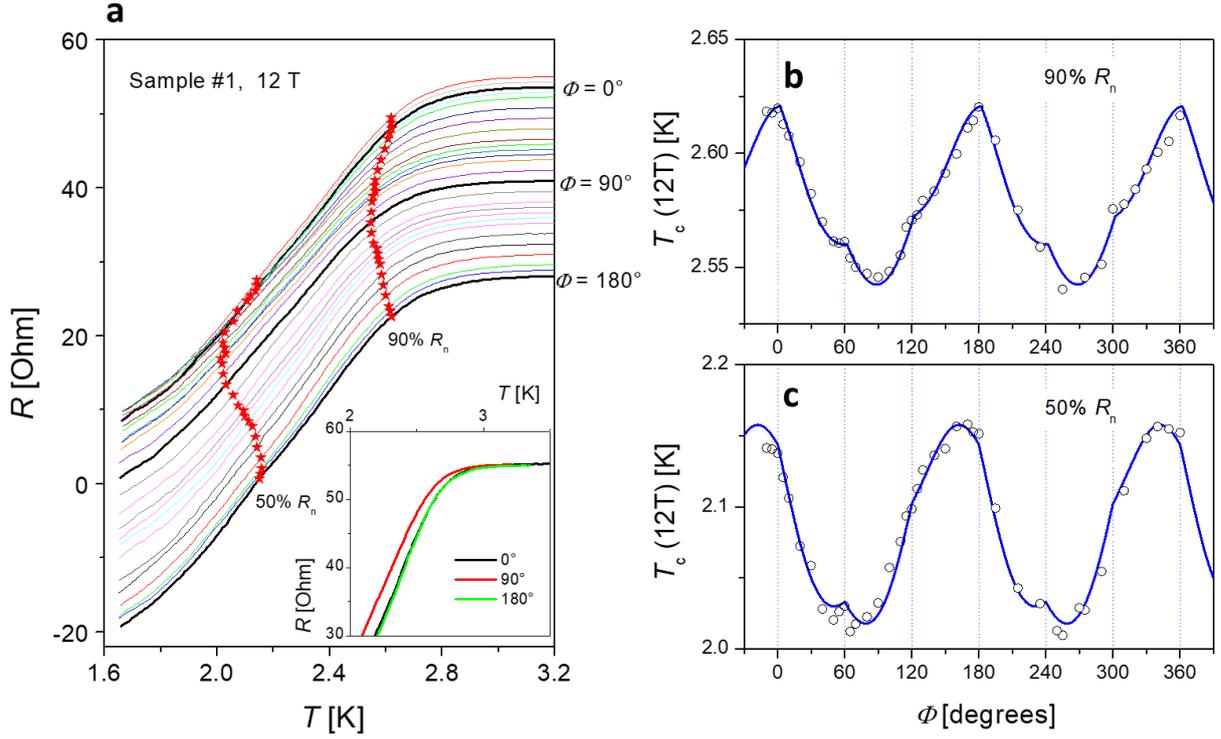

FIG. 2. a) Resistance of Sample #1 in a 12 T magnetic field of different orientations in the plane defined by the angle $\phi$ = -10, -5, 0, 5, 10, 15, 20, 30, 40, 50, 55, 60, 65, 70, 80, 90, 100, 110, 115, 120, 125, 130, 140, 150, 160, 170, 175, 180°. The data were shifted vertically for clearer presentation, except for the -10° data. The inset shows the 0°, 90° and 180° data without offset to illustrate how the curve shifts to lower temperature at 90°. The stars mark temperatures at which 50%, and 90% of the normal state resistance $R_N$ is reached. (b,c) In-plane field-angle dependence of the two characteristic temperatures marked in (a), which illustrates the symmetries of the resistance data in the plane in two different temperature ranges. The lines represent fitting functions considering a 6-fold nodal symmetry together with a two-fold nematic symmetry (see text for details).

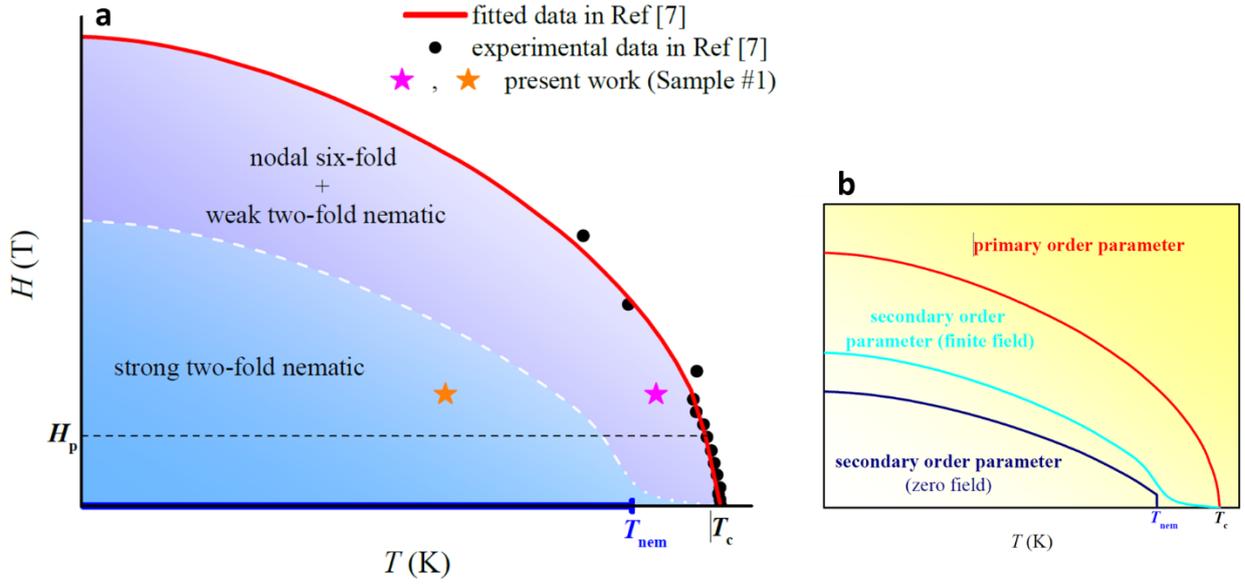

FIG. 3. a) Assumed magnetic field vs. temperature phase diagram. The horizontal line marks the Pauli limit $H_P$. The star marks the regions in which we observe evidence of a six-fold nodal (red) or two-fold nematic order parameter symmetry, respectively. b) Temperature dependence of the primary and secondary order parameter attributed to the nematic superconducting state (see text for details).

# Supplementary Information

**Distinct nodal and nematic superconducting phases in the 2D Ising superconductor NbSe$_2$**

by Cho et al.

|  | $T_c$ (0.5$R_N$) | Number of layers |
|---|---|---|
| **Sample #1** | 3.0 K | Monolayer |
| **Sample #2** | 4.4 K | Monolayer |
| **Sample #3** | 3.7 K | Mono / bilayer |
| **Sample #4** | 4.7 K(2), 5.9 K(3), 7.2 K(Bulk) | Mixture (bilayer + 3-layer + bulk) |

**Supplementary Table 1** | Summary of the 2D NbSe$_2$ devices used in this study. The critical temperature $T_c$ is determined by the midpoint of the resistive transition (0.5$R_N$).

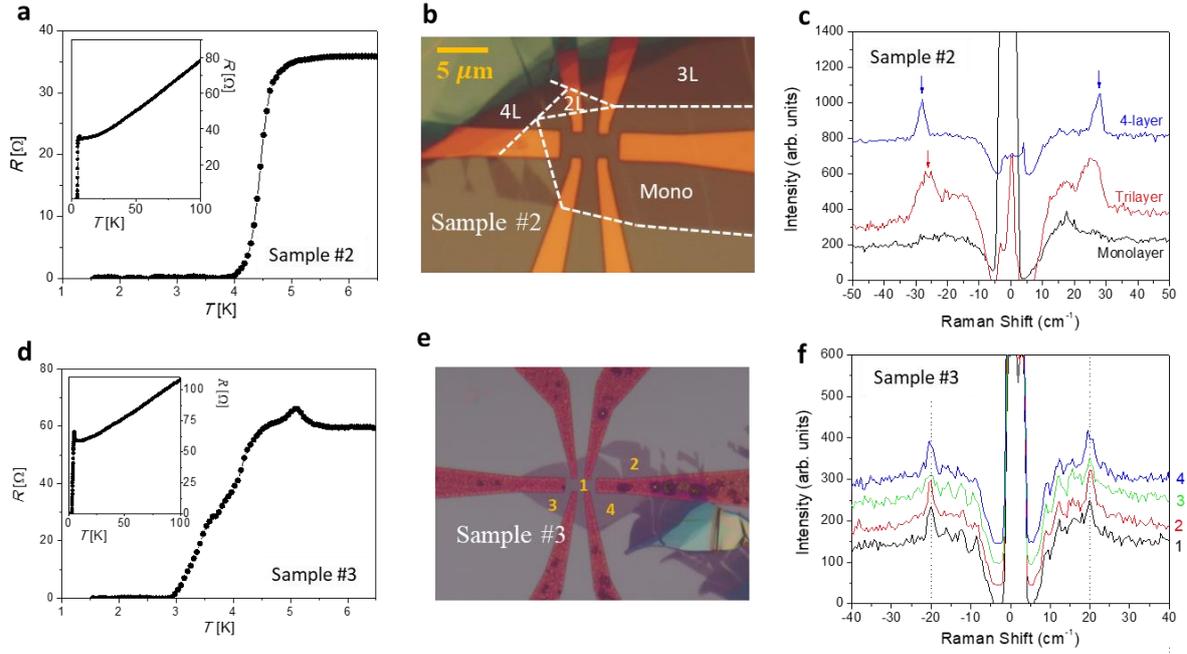

**Supplementary Figure 1 | Device characterization of Sample #2 and Sample #3. (a & d)** Zero-field electrical resistance of Sample #2 and Sample #3, respectively. Sample #2 shows a sharp superconducting transition at 4.4 K, while Sample #3 exhibits a characteristic double transition attributed to mixed mono and bilayer behavior **(b & e)** Optical image of the devices and **(c & f)** local Raman spectra at room temperature recorded at different positions as indicated in the optical images **(b)**. The Raman spectra in (**c**) proves the large monolayer region between the electrodes in Sample #2. The Raman spectra in (**f**) provide only evidence of a bilayer in Sample #3, while the double transition in resistance in (**b**) suggests that the two layers are partially separated or loosely bonded. For clarity, the spectra were shifted vertically.

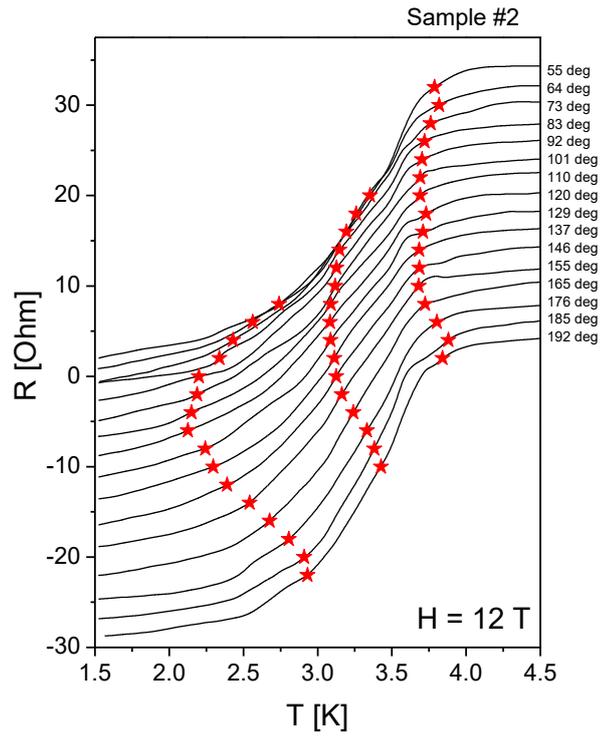

**Supplementary Figure 2 | Field-angle dependence of the resistance of Sample #2.** Resistance of Sample #2 in a magnetic field of 12 T for different field orientations in the plane defined by the angle $\phi$. The data were shifted vertically for clearer presentation, except for the 55° data. The stars mark temperatures at which 25%, 60% and 95% of the normal state resistance is reached at the lower onset, the midpoint and the upper onset of the superconducting transition, respectively, which are included in Fig. 3a-c in the main text.

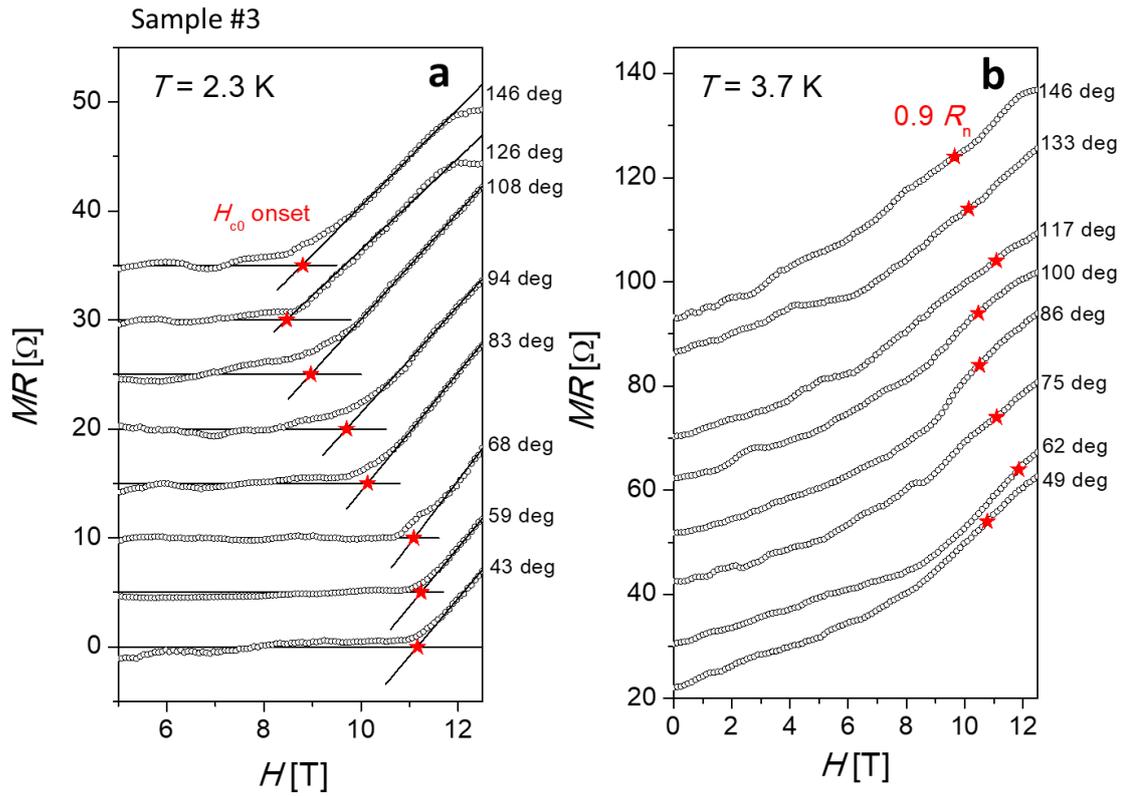

**Supplementary Figure 3 | Field-angle dependence of the magnetoresistance of Sample #3 at two characteristic temperatures.** (**a**) Representative magnetoresistance data at 2.3 K for various field orientations in the plane defined by the angle $\phi$. The stars mark the characteristic fields $H_{c2\,onset}$, which are determined by the intersection of two linear fits, which are included in Fig. 3d in the main text. (**b**) Representative magnetoresistance data at 3.7 K for various field orientations $\phi$ in the plane. The stars mark the characteristic fields $H_{c2}$, which are determined from the field at which the resistance reaches 90% of the normal state value, which are included in Fig. 3e in the main text.

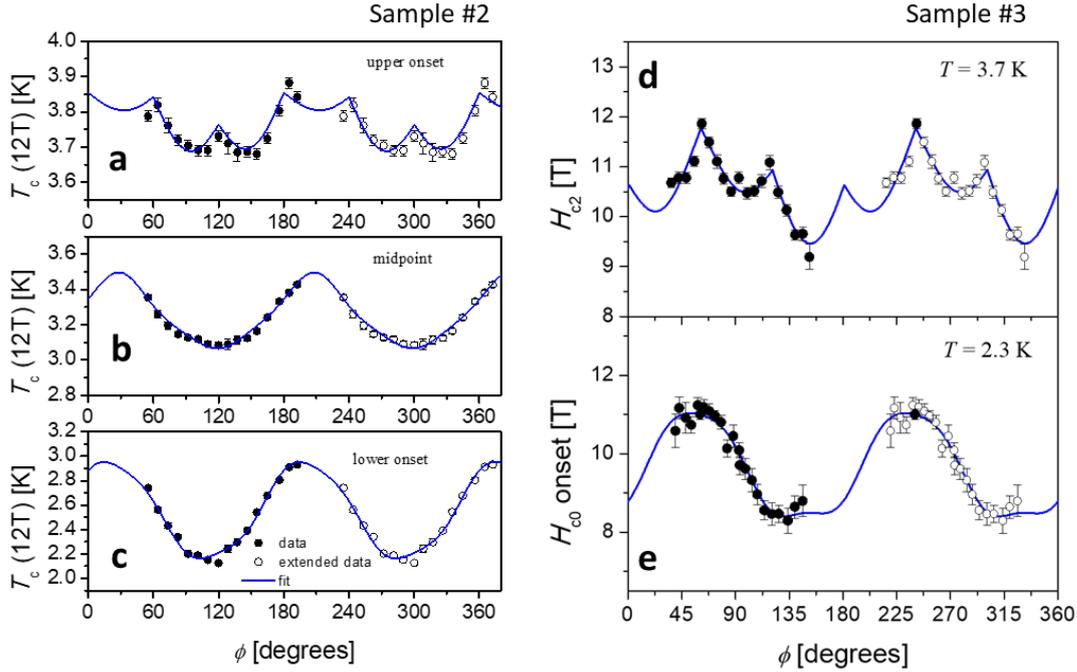

**Supplementary Figure 4** | (a,b,c) In-plane field-angle dependence of Sample #2 of three characteristic temperatures (See Suppl. Fig. 2), which illustrates the symmetries in the plane of the resistance data in three different temperature ranges. The empty circles correspond to the measured data points (represented as filled circles) with a phase of 180 degrees added to it to illustrate the full angular dependence. The blue lines represent fitting functions considering a 6-fold nodal symmetry (a), a 6-fold sinusoidal symmetry (b,c) and two-fold nematic symmetries (a,b,c) (see text for details). (d) Field-angle dependence of the upper critical field $H_{c2}$ obtained from magnetoresistance data of Sample #3 at 3.7 K (see Suppl. Fig. 3b). As criterion we used here when the magnetoresistance reached 90% of the normal state value (note that similar plots can be derived for other criteria as shown in Suppl. Fig. 3). The blue lines represent fitting functions considering a 6-fold nodal symmetry. (e) Field-angle dependence of the characteristics critical field $H_{c0}$ onset, above which zero resistance is lost obtained from magnetoresistance data of Sample #2 at 2.3 K (see Suppl. Fig. 3a). The blue lines represent fitting functions which take into account a weak 6-fold sinusoidal variation in combination with a strong two-fold nematic symmetry. The error bars are determined by the noise level in the resistance data.

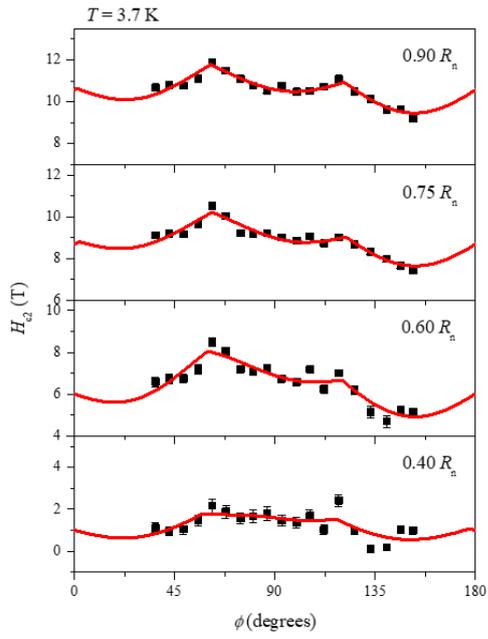

**Supplementary Figure 5 |** Field-angle dependence of the upper critical field of Sample #3 at 3.7 K, derived according to various criteria by selecting characteristics fields where the magnetoresistance reaches 90%, 75%, 60% and 40% of the normal state resistance.

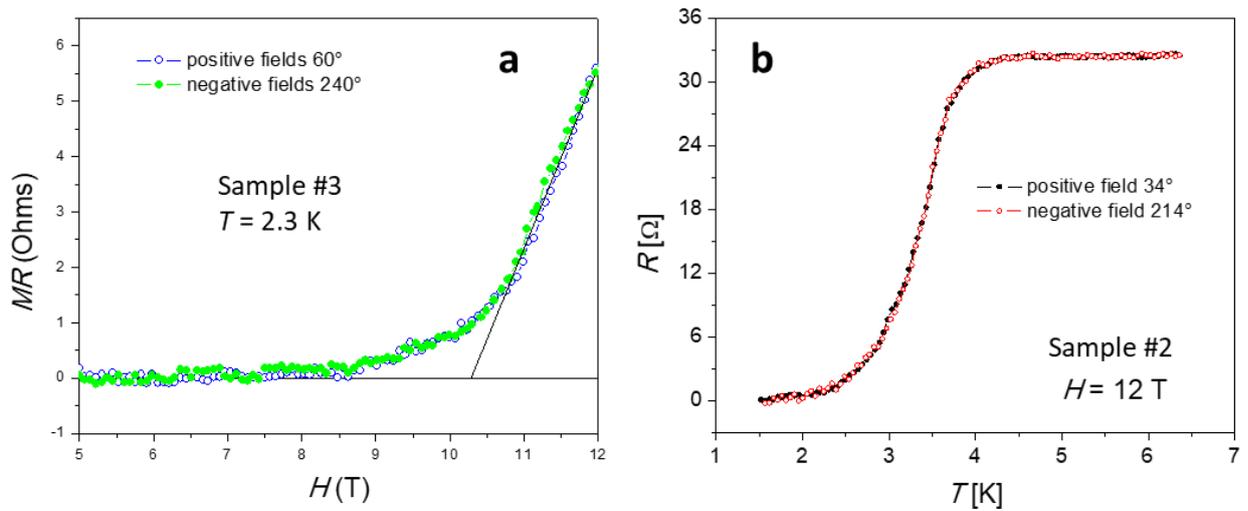

**Supplementary Figure 6 | (a)** Example of magnetoresistance (*MR*) data as a function of the absolute value of the applied magnetic field measured for Sample #3 for two opposite field directions (60° and 240°) by reversing the direction of the magnetic field generated by the superconducting solenoid. The directions of the magnetic field in (a) and (b) were chosen near the maximum upper critical field. The data for the opposite field directions are superimposed and show that time inversion symmetry is preserved and the upper critical field has a two-fold symmetry. **(b)** Example of resistance measured for Sample #2 for two opposite field directions (34° and 214°).

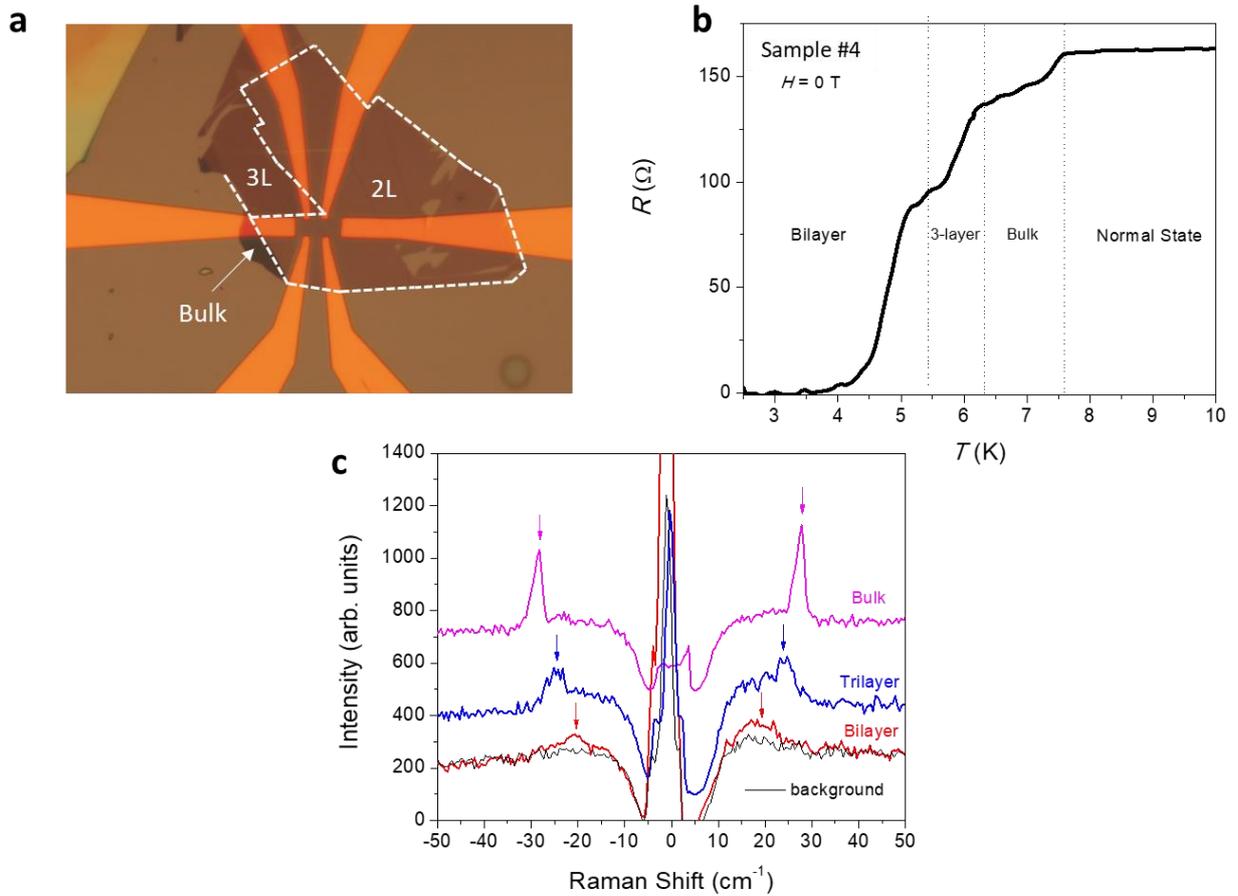

**Supplementary Figure 7 | (a)** Optical image of Sample #4. Three different areas with different thickness can be seen, marked as '2L' (bilayer), '3L' (trilayer) and 'bulk'. The latter refers to a thicker layer in which the superconducting transition temperature and the Raman characteristics approach the characteristics of a bulk sample. **(b)** Zero field electrical resistance as a function of temperature. Each step corresponds to the superconducting transition of an area of different thickness: bulk, trilayer and bilayer. **(c)** Raman spectra recorded on the three regions of different thickness, confirming their layer numbers. The background data were recorded on a bare silicon substrate without $NbSe_2$.

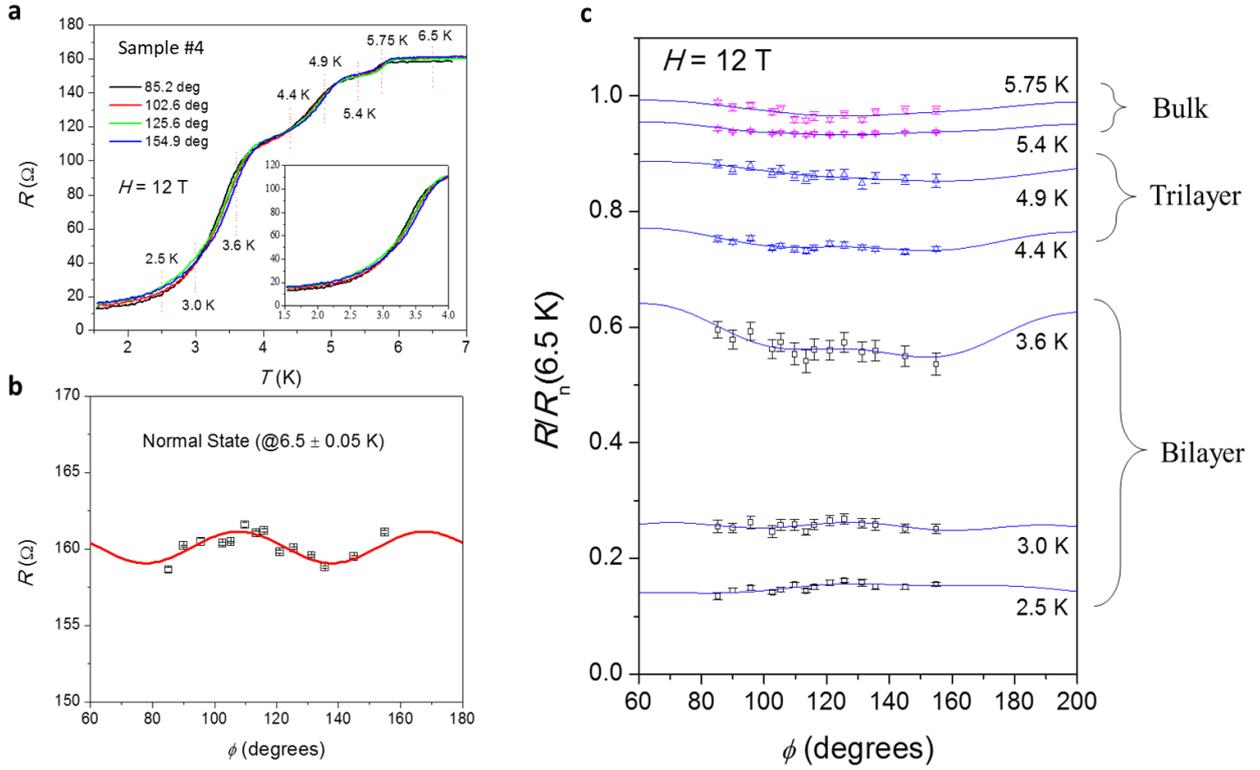

**Supplementary Figure 8 | (a)** Representative field-angle resistance data at a fixed magnetic field of 12 T for Sample #4. **(b)** Field-angle dependence of the resistance at 12 T in the normal state (6.5 K). A weak six-fold symmetry is visible, which was not obvious for Sample #1-3 and is therefore attributed to the bulk $NbSe_2$ region. **(c)** Detailed field angle dependence of the normalized resistance, where we distinguish three different regimes based on each step-like resistive transition. No significant angular dependence can be observed in the bulk superconducting phase. For the temperature regimes dominated by the superconducting transitions of the tri- and bilayer some weak angular variations occur, which can be fitted with a combination of a two-fold and 6-fold node-less symmetry. However, all data of Sample #4 show a significantly lower anisotropy compared to Sample #1-3.

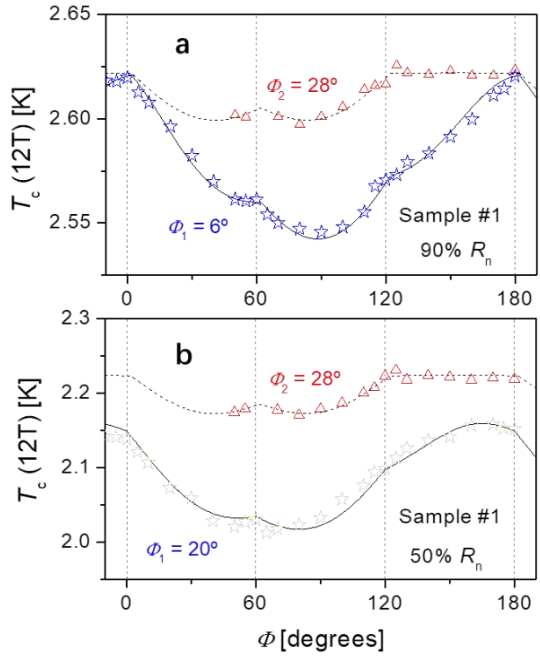
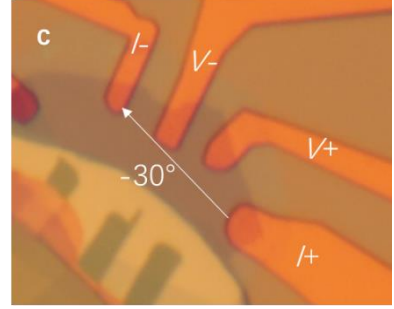
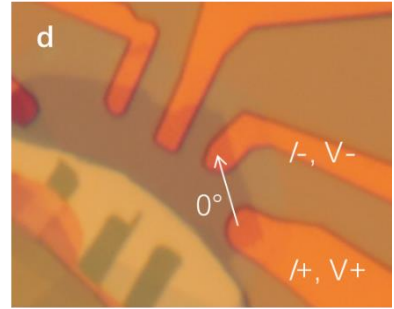

**Supplementary Figure 9 | In-plane field-angle dependence of characteristic temperatures in the resistance of Sample #1 probing different regions of the sample. (a)** In-plane field angle dependence of the temperature at which the resistance reaches 90% of the normal state resistance $R_N$ near the upper onset of the resistive superconducting transition. Stars were measured using a 4-probe technique (see **c** for the configuration of the electric terminals) and triangles were measured using a 2-probe technique (see **d**). Similar data as in **(a)**, but for 50% of $R_N$, which is the midpoint of the resistive superconducting transition. The lines represent fitting functions that account for a 6-fold nodal symmetry together with a two-fold nematic symmetry (see text for details). Note that the orientation of the twofold symmetry for the 2-probe measurement remains along $\Phi_2 = 28°$, while for the 4-probe data (as shown in Fig. 2 of the main text) the orientation varies slightly from $\Phi_1 = 6°$ at 0.9 $R_N$ to $\Phi_1 = 20°$ at 0.5 $R_N$.

# Supplementary Notes

**Overview of the different devices used in this study**

Supplementary Table 1 provides an overview of the devices used in this study. Sample #1 & 2 are dominated by a monolayer region, with the electrodes arranged in such a way that only this region is probed in the magneto-transport experiments. Sample #3 was found to be a mixture of a mono- and a bilayer, while Sample #4 contains bi-layer, tri-layer and thicker regions, all of which show up as a cascade of superconducting transitions due to the different critical temperatures of the regions with different layer thickness. The latter allowed us to probe the field-angle dependence of thicker samples.

**Detailed presentation of results of Sample #2 & 3 (monolayer samples)**

Supplementary Figure 1 shows the device characterization for Sample #2 & #3. Sample #2 shows a sharp superconducting transition at 4.4 K (Supplementary Fig. 1a). The micro-Raman characterization in combination with the optical image shows that the device is dominated by a large monolayer region between the terminals (Supplementary Fig. 1b & c). Sample #3 is a mixture of a monolayer and a bilayer, as shown by the slightly broadened two-stage superconducting transition, which corresponds to the critical temperatures of a monolayer and a bilayer [6], respectively (Supplementary Fig. 1d). The device was particularly small and the local Raman spectra (Supplementary Fig. 1f in combination with the optical image in Supplementary Fig. 1e) were not able to provide separate indications of the different regions.

We measured the resistance in an applied magnetic field of 12 T for Sample #2 and the magnetoresistance for Sample #3 (Supplementary Fig. 2 & 3). For the latter, we obtained better results in measuring the magnetoresistance at fixed temperatures as a function of the magnetic field, allowing us to determine the upper critical field $H_{c2}$ for various directions of the field in the basal plane.

Supplementary Fig. 2 shows resistance data of Sample #2 measured in an applied magnetic field of 12 T for selected in-plane directions of the magnetic field in the plane. Similar to Sample #1, a significant angular variation is observed. We marked characteristic temperatures at which the resistance reaches a certain value near the lower onset, the midpoint and the upper onset of the superconducting transition through stars, which were used to compile the plot of these characteristic temperatures as a function of the angular orientation of the field in the plane in Supplementary Fig. 4a - c. The trend is very similar to that of Sample #1. Here, only the filled circles represent measurement data. The open circles were derived from the same data, but with an angle of 180º added to illustrate the full angular dependence. For some selected angles, we tested that the inversion of the field leads to the same critical fields and thus excludes a broken time inversion symmetry (see Suppl. Fig. 6 and the Supplementary Discussion). At the upper onset, just before the sample enters the normal state, the angular dependence shows a pronounced six-fold variation with sharp spike-like maxima at 60º, 120º and 180º, superimposed by a weaker two-fold variation. At lower temperatures (midpoint and lower transition onset), the spikes disappear, and the data show a pronounced two-fold symmetry.

The magnetoresistance data of Sample #3 in Supplementary Fig. 3 was taken at two different fixed temperatures (2.3 and 3.7 K) for selected field orientations in the plane. It should be noted that Sample #3 is a mixture of a monolayer and a bilayer, but the bilayer has a lower anisotropy than the monolayer so any observed anisotropy is attributed to the monolayer, which is further confirmed by comparison with the data of Sample #1 & 2 (pure monolayers) and #4 (bilayer, trilayer and bulk). At 3.7 K (a) and 2.3 K (b) a significant field-angle variation can be observed in the high magnetic field regime. From both data sets we derive characteristic fields related to the upper critical field $H_{c2}$, which were used to compile the plot of these characteristic fields as a function of field-angle in in Supplementary Fig. 4d & e. At a lower temperature of 2.3 K (lower panel), the onset point ($H_{c0\ onset}$) of the broad upper critical field transition is determined from the zero crossing point of linear fits in high fields where zero resistance is lost (Supplementary Fig. 3a). The magnetoresistance in Suppl. Fig. 3a shows a broadened step between the zero-resistance state and the normal state, and this step is basically shifted to higher or lower fields depending on the field orientation in the plane, while essentially retaining its shape. It thus reflects the angular dependence of the upper critical field in the basal plane. At the higher temperature of 3.7 K (top panel), we used another criterion to define the upper critical field $H_{c2}$, because at this temperature we are already in the middle of the resistive transition of the monolayer where the zero resistance is lost. Thus, $H_{c2}$ was used to see the angle-dependent variation defined as the field in which 90% of the normal state resistance (0.9 $R_N$) is restored. At 3.7 K, the angular dependence of $H_{c2}$ (Supplementary Fig. 4d) shows very similar peaks with a 6-fold rotational symmetry as the critical temperature of Sample #1 & 2, while at 2.3 K (Supplementary Fig. 4e) the characteristic field $H_{c0}$ onset, above which the zero resistance is lost (Suppl. Fig. 3b), shows a pronounced 2-fold symmetry. In Supplementary Fig. 5 we show the field-angle dependence of $H_{c2}$ measured at 3.7 K using various criteria to define the upper critical field from the broadened magnetoresistive transition using different criteria by selecting characteristics fields at which the magnetoresistance reaches 90%, 75%, 60% and 40% of the normal state resistance. The kink-like structure can be clearly resolved at least down to fields where 60% of the normal state is reached.

The behavior of all three samples (#1 - #3) is very similar, which demonstrates the reproducibility of our results in 3 separate devices.

**Investigation of the preservation of the time inversion symmetry**

While we achieved to cover a full angular range of more than 360° in our experiment on Sample #1, which shows identical results if the field direction is reversed, the angular range of our experiments for Sample #2 - 4 was limited for technical reasons. This angular range is sufficient to study the 6-fold symmetry of the nodal structure near the upper critical field line. Assuming that time inversion symmetry is valid, i. e. that a 180° reversal of the magnetic field direction leads to identical results, the angular range is large enough to identify the nematic superconducting state as well, but this is not necessarily self-evident, since e.g. a chiral triplet superconducting state would break the time inversion symmetry. The investigation of the additional two-fold nematic symmetry therefore required a further test to be able to distinguish a possible two-fold nematic symmetry from a three-fold time-inversion breaking symmetry.

For Sample #3 a maximum of the field-angle dependence was observed at 55° (Suppl. Fig. 3d). In case of a two-fold symmetry, a further maximum would be expected when the field direction is reversed to 235°, so that the time-reversal symmetry is preserved. On the other hand, in the case of a three-fold symmetry a minimum at 235° would be expected when the field direction is reversed, indicating a broken time-inversion symmetry. In Suppl. Fig. 6a we show magnetoresistance data measured for 'positive' and 'negative' field directions by reversing the current und thus the field direction in our bipolar superconducting solenoid. It can be seen that both data are perfectly superimposed, which demonstrates that the time inversion symmetry is maintained. The additional points measured in this way are included in Suppl. Fig 4d. For Sample #2 we show resistance data measured in +12 T and -12T in Suppl. Fig. 5b corresponding to the directions 34° and 214° in the plane. Here again, the two resistance curves merge perfectly, extending our angular range and confirming that the time inversion symmetry is maintained. The additional points obtained in this way have been added to Fig. 4a-e. This is in perfect agreement with the measurements on Sample #1 where we covered the full angular range of 360°.

Therefore, our field-angle dependence can only be described by combinations of 6- and 2-fold symmetries, and superconductivity preserves the time inversion symmetry, so that other exotic, e.g. chiral, superconducting states are excluded.

**Detailed presentation of results from Sample #4**

Sample #4 contains contributions from different layer thicknesses. The characterization of the device is shown in Supplementary Fig. 7. White lines in the optical image in Supplementary Fig. 7a illustrate the different regions of different thickness, which were identified with help of the local micro-Raman spectra (Supplementary Fig. 7c). The main regions between the six contacts is represented by a bi-layer, but also a tri-layer region is extending in between the two upper voltage probes and thus will be probed as a parallel resistance. In addition, there is a thicker region which is marked as 'bulk'. Supplementary Fig. 7b shows the zero field resistance of the sample in which a cascade of at least three superconducting transitions is visible. The lowest two steps occur at the characteristic temperatures known from bi-layer and tri-layer $NbSe_2$, respectively [7], and can be attributed to the corresponding regions marked in the optical image. The upper one occurs at the same temperature as $T_c$ in bulk samples and is attributed to the third region.

In Supplementary Fig. 8 we show the field-angle dependence data for Sample #4. Supplementary Fig. 8a shows the electric resistance for 4 selected orientations with respect to the in-plane direction of a 12 T magnetic field applied parallel to the layers. While there is a certain very weak dependence on the in-plane field angle, it is obvious that the angular variation is much weaker than for the monolayer devices Sample #1 - #3. In Supplementary Fig. 8a & b we plot the resistance as a function of the in-plane direction of the magnetic field at various characteristic fixed temperatures. The field-angle dependence of the bilayer region is significantly weaker than that of Samples #1, 2 & 3 and is dominated by a weak six-fold variation (Suppl. Fig. 8c). At the center of the partial superconducting transition at 4.4 K, an even weaker anisotropic structure is formed, which we attribute to the superconducting transition of the trilayer region in the device. Partial superconducting transitions, which occur at even higher temperatures and are assigned to thicker sample regions, show an almost isotropic behavior, as is expected for the bulk limit. It is hard to draw any clear conclusion about the field-angle symmetry in any of the different temperature regions, since the field-angular variation is too weak to be significant. The pronounced nodal and nematic symmetries appear thus to be a special characteristics of the monolayer case, unless there is a certain domain structure in Sample #4 in which domains with different nematic symmetries are oriented along different crystalline directions.

**Evidence of multiple nematic domains in Sample 1 #**

In Suppl. Fig. 9, we compare the in-plane field-angle dependence of characteristic temperatures in the superconducting transition of Sample #1, at which the resistance reaches 90% of the normal state resistance (a) and 50% of the normal state resistance (b), probing two different regions of the $NbSe_2$ monolayer. We use the same 4-probe data as shown in Fig. 2 of the main text to probe the region between the two voltage terminals (see Suppl. Fig. 9c for the configuration of the electric terminals), and we use a 2-probe technique to investigate another region near the edge of the monolayer (Suppl. Fig. 9d). The lines represent fitting functions that consider a 6-fold nodal symmetry together with a two-fold nematic symmetry (see main text for details). From the fitting parameters, the orientation of the twofold symmetry for the 2-probe measurement remains along $\Phi_2 = 28°$, while for the 4-probe data the orientation varies slightly from $\Phi_1 = 6°$ at 0.9 $R_N$ to $\Phi_1 = 20°$ at 0.5 $R_N$. This indicates the presence of multiple nematic domains, which is not uncommon in nematic superconductors.

Note that the current direction for the 4-probe technique is approximately along -30°, while for the 2-probe measurement it is along 0°, which does not seem to show any correlation with the orientations of the 2-fold symmetry.

**Discussion of the effect of the near degeneracy of the two order parameters in bulk NbSe$_2$**

A question is whether the near degeneracy of the two order parameters.is also present in bulk NbSe$_2$ samples and if so, why it does not induce a nematic state as well. An answer to the second question might be that the near degenerate pairing state in bulk NbSe$_2$ is a *p*-wave triplet state with $E_{1u}$ symmetry. For such a state no coupling of the type given in Eq. 4 is symmetry allowed and therefore no nematicity is expected. Hence, one of the implications of our findings is that bulk NbSe$_2$ should have a dominant *s*-wave order parameter with a potential sub-leading *p*-wave triplet channel, a prediction that could be probed by optical excitation of the sub-leading order in analogy to the Bardasis Schrieffer mechanism [4].

**Comparison of the field dependence of our samples with literature data**

There is little magnetoresistance data for monolayer NbSe$_2$ for in-plane magnetic fields, and while in Ref. [1] there is hardly any field dependence up to 9 T, there is a field dependence similar to our case in Ref. [2]. Unlike most conventional bulk superconductors, monolayer samples are inherently very delicate, and it is not surprising that their properties are dramatically affected by impurities, different types of substrates and capping layers, as well as strain induced by our transfer process. Our applied field of 12 T is already ~1.6 times higher than the Pauli limit, and in general, one can see in the phase diagrams of Ising superconductors in the literature that the slope of the critical field line gets reduced not too far beyond the Pauli limit, and thus $T_c(H)$ is typically suppressed faster. In Ref. [3] it was shown that the complex interplay between singlet and triplet channels with scalar impurities has a strong effect on the superconducting transition at a finite magnetic field and thus can lead to strong variations of the slope of the upper critical field transition line, especially above the Pauli limit, since the scattering rate "undoes the enhancement caused by SOC" [3].

**Supplementary References**